# Localism as Secrecy: Efficiency-Secrecy Tradeoffs in the Recruitment of ISIS Foreign Fighters

Nate Rosenblatt
University of Oxford
Nuffield College – Department of Sociology

**Abstract**

This paper compares networks of foreign fighters who joined the Islamic State of Iraq and Syria (ISIS) from Europe and the Arabian Peninsula in order to test whether there are differences in their recruitment and how those differences affect the nature of the foreign fighter mobilization. It is the first study to compare different networks of foreign fighters that joined the same group in the same conflict at the same period of time.

This study finds that foreign fighter recruitment resembles an efficiency-secrecy tradeoff: in places where recruitment needs to be hidden from legal scrutiny, recruitment networks are decentralized; composed of small and more local recruitment cells. These cells can operate more secretly and the group as a whole is more resilient to disruption. In exchange, it is hard for the group to attract large numbers of recruits. Whereas in places where recruitment could occur more freely, recruitment networks are more hierarchical; comprised of a larger number of recruits with more geographically diverse connections. The hierarchical design of their recruitment networks may be easier to disrupt, but it also helps the group efficiently recruit more followers if left undisturbed. This study concludes that the ISIS foreign fighter recruitment process varied significantly. Researchers and policymakers focused on recruitment and radicalization should therefore carefully frame their results or policies based on the different types of recruitment processes and the various social, political, and legal contexts where their work takes place.

## Table of Contents

## I. Introduction

Two years into the Syrian civil war, in the spring of 2013, a group of militant commanders met for five days in a village called Kafr Hamra on the outskirts of the city of Aleppo, Syria. Over grilled chicken, french fries and tea, Abu Bakr al-Baghdadi, then-commander of the Al-Qaeda-linked Islamic State of Iraq, tried to persuade his audience to help him create a physical Islamic state. He argued that borders, citizens, institutions, and a functioning bureaucracy would offer a home to Muslims from all over the world. Unlike the elusive Al Qaeda (AQ), he explained, a formal state would attract recruits from among ordinary Muslims who might otherwise not know how to join.[1] In April 2013, weeks later, he announced the group, once called the Islamic State of Iraq, would be rebranded into the Islamic State of Iraq and Syria (ISIS).[2]

Al-Baghdadi's prediction about the allure of an Islamic state for new recruits proved prescient. Estimates of ISIS' foreign fighters describe the group as having at least 40,000 fighters from over 100 countries foreign to Iraq and Syria.[3] This was not only by far the largest mobilization of foreign fighters for any insurgent group in the Syrian war, it was perhaps the largest mobilization of foreign fighters to any conflict since at least the Spanish Civil War.[4]

This study will focus on how those foreign fighters were recruited to join ISIS in Syria and Iraq. It conducts a social network analysis of foreign fighters from different countries in Europe and the Middle East and North Africa region (MENA) to understand the extent to which recruitment pathways into the group vary by region. With the data currently available, it is the first opportunity to conduct a comparative study of how foreign fighter recruitment networks varied by region, even when these fighters joined the same group in the same conflict at the same time.

---

[1] Harald Doornbos and Jenan Moussa, "Present at the Creation: The Meeting that Led to the Beginning of ISIS," *Foreign Policy*, August 16, 2016, http://foreignpolicy.com/2016/08/16/present-at-the-creation/ (accessed January 2, 2019).

[2] I retain the original name of the group – the "Islamic State of Iraq and Syria" (ISIS) in this paper because it was not renamed the "Islamic State" until July 2014, which was mostly beyond the scope of the time period this paper analyzes.

[3] Richard Barrett, "Beyond the Caliphate: Foreign Fighters and the Threat of Returnees," *The Soufan Group*, November, 2017, http://thesoufancenter.org/wp-content/uploads/2017/11/Beyond-the-Caliphate-Foreign-Fighters-and-the-Threat-of-Returnees-TSC-Report-October-2017-v3.pdf (accessed December 14, 2018).

[4] I also use Malet's comparative historical research to define foreign fighters as "non-citizens of a state experiencing civil conflict who arrives from an external state to join an insurgency." Malet, David, "Foreign Fighter Mobilization and Persistence in a Global Context," *Terrorism and Political Violence* 1 (2015): 6; See also: David Malet, *Foreign Fighters: Transnational Identity in Civil Conflicts* (Oxford: Oxford University Press, 2013).

It compares networks of ISIS foreign fighters who joined in 2013-14 from Europe and the Arabian Peninsula in order to demonstrate how structural differences exist in the ISIS foreign fighter recruitment process.[5] This paper will be divided as follows. The next section (section II) will review the study's methodology, including how I obtained, cleaned and presented the data in a format for social network analysis. Section III compares basic network statistics between fighters from Europe and the Arabian Peninsula, section IV analyzes the background characteristics of these fighters (i.e., marital status, age, educational attainment, professional history, their home country/region) that best predicts whether they know each other another using a type of mathematical modeling for social networks called exponential random graph models (ERGMs). Section V will discuss the mechanisms behind these results, and section VI will conclude by summarizing 1) what these tests add to our understanding of how foreign fighters are recruited, 2) the limitations of this approach, and 3) the implications of this paper for future research, particularly that which compares online and offline recruitment.

### *A Theory of Localism as Secrecy*

Localism – the preference for one's own area or region – served as a stand-in for secrecy of the foreign fighter recruitment effort. The more secrecy was required in recruitment, the more likely recruits would have fewer ties, more local ties, and the structure of their network would be decentralized. This study concludes that a fighters' geographic origin – country or province – is the single most significant attribute predicting ties with other fighters. It is a better predictor than any other individual-level variable (e.g., marital status, age, profession, or previous fighting experience) in determining whether two fighters know each other.

[5] The Arabian Peninsula states include: Kuwait, Bahrain, Qatar, the United Arab Emirates, Oman, Yemen, and Saudi Arabia. Europe includes all participants in the Schengen Area, the United Kingdom, and Balkan states.

On the one hand, this seems like a fairly obvious finding – of course it is more likely that a fighter from Belgium will be connected to other Belgians than he would be to someone from Saudi Arabia. The finding is more useful when framed in *comparative terms*: the question at the heart of this study is not whether Belgians are more likely to be connected to other Belgians, but *how much more likely* are Belgians connected to other Belgians than Saudis are to other Saudis?

This paper finds that the predictive strength of geographic origins on tie formation between ISIS foreign fighters varies considerably depending on where a foreign fighter is originally from. It finds that geography is far more predictive of a tie between fighters from Europe than for those from the Arabian Peninsula. In Europe, fighters who shared a country of origin were about 11 times more likely to share a tie than fighters who shared a country of origin in the Arabian Peninsula. Fighters who shared a province level tie in a European country were about five times more likely to be connected to each other than fighters who came from a country in the Arabian Peninsula. That means there was a far greater restriction to the geography of ties among fighters in Europe than of those from the Arabian Peninsula.

I believe the reason why the networks of fighters from the Arabian Peninsula were larger, stronger, more hierarchical, and more geographically dispersed than those from Europe is because of how free they were to self-organize and mobilize. Anyone interested in joining ISIS from the Arabian Peninsula could meet like-minded individuals in any other country or province in the region, virtually unimpeded.[6] However, in Europe, there were steeper barriers to transnational ties. This included pervasive social and legal opposition.[7] The scrutiny of Europe's socio-legal system on these networks is likely why the geographic spread of ISIS foreign fighter ties in Europe was far more limited than in the Arabian Peninsula.

---

[6] See, for example, Dickinson, Beth. (2013, December). Playing with Fire: Why Private Gulf Financing for Syria's Extremist Rebels Risks Igniting Sectarian Conflict at Home. Brookings Institute. Analysis Paper No. 16. Retrieved from https://www.brookings.edu/wp-content/uploads/2016/06/private-gulf-financing-syria-extremist-rebels-sectarian-conflict-dickinson.pdfal-.
[7] See, for example, Olivier Roy, *Jihad and Death: The Global Appeal of Islamic State* (Oxford: Oxford University Press, 2017).

In *The Terrorist's Dilemma*, Jacob Shapiro described how terrorist groups face a tradeoff between efficiency and security in their operations.[8] They would prefer to develop organizational charts, clear hierarchies and lines of communication. This would allow them to function more efficiently – the leaders of the group could issue orders and have clear mechanisms to monitor whether those orders were being carried out. But, of course, this is rarely possible, so terrorist groups design decentralized organizational structures to maximize secrecy and minimize the effect of disruption. In a decentralized network, getting rid of one cell or leader does not dismantle the whole group.

But Shapiro's study focused on how a terrorist group behaves. This study appears to confirm that there is a secrecy-efficiency tradeoff not just in terrorist operations, but also in terrorist group recruitment. This tradeoff is found in the *localism* of the recruitment process. In the Arabian Peninsula, prospective fighters were able to meet other like-minded individuals from across the country without nearly as much scrutiny from law enforcement as in Europe. In Europe, prospective fighters were recruited with concerns for law enforcement. Therefore, fighters from Europe made fewer ties, and the connections they made were highly local. Consequently, their networks were also comprised of a smaller number of recruits.

Previous work on terrorist group recruitment has underscored the primacy of social networks. Julie Hwang found that social network ties were the most important dimension facilitating the recruitment of Indonesian foreign fighters.[9] Marc Sageman studied 175 individuals who joined Al Qaeda (AQ) and found that 75% of them joined with at least one other person.[10]

---

[8] Jacob N. Shapiro, *The Terrorist's Dilemma: Managing Violent Covert Organizations* (Princeton, NJ: Princeton University Press, 2013).
[9] Chernov Hwang, J. (2018), "Pathways into Terrorism: Joining Islamist Extremist Groups in Asia," a special issue of *Terrorism and Political Violence*.
[10] Marc Sageman, *Understanding Terror Networks* (Philadelphia: University of Pennsylvania Press, 2004).

Another multi-nation study concluded that 95% of AQ fighters were recruited by a friend or family member.[11] A study of Dutch foreign fighters from 2000-2013 found that "social facilitators" were a key step in the transition from radicalization to mobilization.[12] In Turkey, a study of foreign fighters found that 96% of fighters cited "peer pressure," and 75% cited "socialization with other foreign fighters" as part of their decision-making process to fight in Syria.[13] In Germany, Reynolds and Hafez study developed detailed profiles for 99 German foreign fighters, finding that "71 of the 99 profiles had a personal connection with at least one other German foreign fighter before departing Germany."[14] The same is true in many other contexts.

This past work emphasizes that networks matter in foreign fighter recruitment, but it lacks comparative context. Are some places more prone to extensive in-person recruitment than others? To address this gap, this paper demonstrates empirically that the networks of foreign fighters vary considerably depending on what region they come from. It also narrows down the factors that best explain these differences and what it means for how foreign fighters join an insurgency. For the first time, we have data that can show us differences in how people can be recruited for the same group in the same conflict at the same time but from different places.

## II. Data and Methods

In early 2016, an ISIS fighter defected with personnel records containing detailed registration forms recorded by ISIS border officials for newly arrived foreign recruits from 2013-

---

[11] Justin Magouirk, Scott Atran, and Marc Sageman, "Connecting Terrorist Networks," *Studies in Conflict & Terrorism* 31, no. 1 (2008): 1-16.
[12] Jasper L. de Bie, Christianne J. de Poot, and Joanne P. van der Leun, "Shifting Modus Operandi of Jihadist Foreign Fighters from the Netherlands between 2000 and 2013: A Crime Script Analysis," *Terrorism and Political Violence* 27, no. 3 (2015): 416–40.
[13] Murat Haner, Ashley Wichern, and Marissa Fleenor, "The Turkish Foreign Fighters and the Dynamics behind Their Flow into Syria and Iraq," *Terrorism and Political Violence* (2018), DOI: 10.1080/09546553.2018.1471398.
[14] Sean C. Reynolds and Mohammed M. Hafez, "Social Network Analysis of German Foreign Fighters in Syria and Iraq," *Terrorism and Political Violence* (2017), DOI: 10.1080/09546553.2016.1272456, p. 16

2014. I acquired these forms through my contacts with Syrian journalists, verified their validity, translated, de-duplicated, and hand-coded them from their initial form – thousands of individual PowerPoint files – into a single spreadsheet totaling 3,581 individuals.[15]

In the forms, recruits were asked to disclose their birthdays, names, home addresses, and other personal details. For a more complete definition of these questions and statistics about their distribution across the entire sample, please see previous studies using these data.[16] A translated list of questions in the form and a copy of the Arabic version can be found in Annex I and II. Table 1 below describes the attributes in the foreign fighter registration forms that I analyzed in this study.

Table 1 - Node Attribute Definitions Considered in This Study

| # | Attribute | Variable Type | Scoring | Description |
|---|---|---|---|---|
| 1 | Country/Region of Residence | String | 'France'/ 'Strasbourg' | The country in which the respondent reports last residence. |
| 2 | Birth Year | Continuous | 19XX | The year that each respondent was born |
| 3 | Nationality | String | 'France' | The self-reported nationality of the respondent. Occasionally, respondents will note their family origins and their passport status (e.g., "Tunisian family, French passport"). This study used the latter if a distinction was made. |
| 4 | Marital Status | String | 'Single' | The marital status of the fighter, which included 'single,' 'married,' or 'with children.' |
| 5 | Education Score | Discrete | 1-5 | The level of educational attainment, ranked from 1 (no education) to 5 (university or higher). |
| 6 | Religious Knowledge | String | 'Elementary' | The level of self-reported religious knowledge, which included three options: 'elementary,' 'moderate,' and 'knowledgeable.' |

[15] Rosenblatt (see note 23 above).
[16] Brian Dodwell, Daniel Milton, and Dan Rassler, "The Caliphate's Global Workforce: An Inside Look at the Islamic State's Foreign Fighter Paper Trail," *Combating Terrorism Center, West Point*, April, 2016, Retrieved from: https://ctc.usma.edu/wp-content/uploads/2016/04/CTC_Caliphates-Global-Workforce-Report.pdf (accessed December 2, 2018).

| | | | | |
|---|---|---|---|---|
| 7 | Profession | Discrete | 1-5 | The fighters' previous job, ranked from 1 (unemployed) to 5 (advanced professional). |
| 8 | Previous Jihad Experience | Dichotomous | 0, 1 | Whether the respondent reported to have fought previously in a "jihad". |
| 9 | Countries Visited Score | Discrete | 1-5 | 1 = Zero countries; 2 = 1 country; 3 = 2 countries; 4 = 3-4 countries; 5 = 5+ countries |
| *For additional detail on the questions included in the form, please see the complete questionnaire in the Annex.* | | | | |

This study will analyze networks of fighters from Europe and the Arabian Peninsula to compare what factors best predict their ties to one another. The hypothesis is that geographic origin is the strongest predictor of ties between fighters ('Attribute 1' in Table 1). The other attributes are mainly listed above as "controls," – which means we can say how much geographic origin mattered for foreign fighters from one region over another regardless of a fighters' age, marital status, professional and educational backgrounds, etc. These other variables also help us understand how naturally forming the networks of foreign fighters are based on the theory of "homophily."[17] Homophily simply states that, absent other influences, people tend to connect with others who share similar backgrounds. In this case, if a fighters' age, marital status, educational and professional backgrounds predict their knowing another fighter, that network is more likely to be naturally forming – or absent any external force (i.e., manipulation by ISIS recruiters or pressure from law enforcement scrutiny).

To specifically study foreign fighters from Europe and the Arabian Peninsula, this study selected a 1,059 out of the 3,581 coded fighters in the entire sample based on two criteria:

(1) The fighter either claimed *residence in* a European country (defined as European Union plus the United Kingdom) or in an Arabian Peninsula country.
(2) The fighter *self-reported travel to* a European country or an Arabian Peninsula country in their form.

[17] M. McPherson, L. Smith-Lovin, and J.M. Cook, "Birds of a Feather: Homophily in Social Networks," *Annual Review of Sociology* 271, no. 1 (2001): 415-44.

The reason why I include people who traveled to Europe or the Arabian Peninsula, rather than just people who report living in that region, is to determine the extent to which fighters have 'transnational' ties. If local ties did not matter, or if they mattered less than other factors, then it is worth testing whether fighters met other fighters who traveled to their region. A total of 1,578 fighters were analyzed in this study: 519 fighters in the Europe sample and 1,059 in the Arabian Peninsula sample (Please refer to the tables in Annex III and IV for a list of the geographic origin of fighters in this study).

Finally, I chose to compare fighters from the Arabian Peninsula and Europe because they are different from each other, and therefore likely to produce dissimilar results. But the regions also have some key similarities, in particular reasonably free trade and cross-border travel during the time they were recruited, as well as historical, political, social and cultural similarities. I selected to compare regions, rather than specific countries, in order to avoid the possibility that differences discovered were a product of the specific conditions in those countries (although in Appendix VI, I conduct the same test for two countries and find similar results).

### *Generating and Analyzing Network Data*

I constructed three different networks for each region, one for each of three different questions on the leaked ISIS foreign fighter registration forms. First, ISIS fighters were asked to name the person who was their "Referral" (*Tazkiyya*) – someone who could vouch for their jihadist "credentials." Second, they were asked to name the "Facilitator" (*Wassita*) – someone who helped smuggle them into ISIS territory. And third, the ISIS administrator noted the date and location of where the newly arrived fighter entered ISIS territory ("Date/Location"). Fighters were connected to each other if they shared the same reference in common, the same facilitator in common, and/or whether they arrived on the same day and at the same location. Definitions for these three networks are provided below in Table 2 below. Fighters are "nodes", the factors determining connections between them are "edges."

Table 2 – Node-Edge Definitions for the Three Networks in This Study

| **Network #** | **Node** | **Edge** | **Network type** |
|---|---|---|---|
| 1 | Fighters | Referrals | Undirected, *weighted* network in which fighters are connected based on whether they mentioned the same referral who recommended that they join. The network is weighted when two fighters share multiple referrals, although this is rare. |
| 2 | Fighters | Facilitator | Undirected, unweighted network in which fighters are connected based on whether they mentioned the same facilitator who helped them reach ISIS territory. |
| 3 | Fighters | Date/Location of Entry into ISIS territory | Undirected, unweighted network in which fighters are connected based on whether they were logged as entering ISIS territory at the same time and place. |

All three of the networks described above are "two-mode networks." Two-mode networks are those that *infer* ties based on some mutual connection, such as co-attendance at social gatherings or co-participation in a protest.[18] The strength of the inference of a tie between two people varies in two-mode networks: if two people attend a protest of 10,000 people, we are less likely to infer that they know one another than if they attended a meeting of 10 people.

Besides size of an event they attend, another way to infer how strong a two-mode network can predict ties between two nodes is if the network demonstrates certain homophily attributes.[19] For example, if the fighters are connected to each other and have nothing in common with one another, there may be some external force ordering their connections (i.e., ISIS recruiters). Or these networks only weakly model actual relationships between fighters (see "Limitations" in Section VI). In other words, if the networks in this study do not display any homophily, we should be suspicious of our how accurately they model reality absent a compelling explanation for some external ordering of the fighters' ties.

---

[18] Stephen P. Borgatti, Martin G. Everett, and Jeffrey C. Johnson, *Analyzing social networks* (London: SAGE Publications Limited, 2013).
[19] Ivan Brugere, Brian Gallagher, Tanya Y. Berger-Wolf, "Network Structure Inference, A Survey: Motivations, Methods, and Applications." *ACM Computing Surveys* 51, no. 2, 24.

I use three different analytical techniques to analyze the networks in this study. First, I use *Gephi,* a network analysis software, to collect and analyze descriptive network statistics. This includes the average number of ties with other fighters each fighter has, and the network's "degree density," or the proportion of actual ties to total possible ones.[20] This approach will tell us which network of fighters, from Europe or the Arabian Peninsula, was more interconnected.

Second, I combine all three networks into a single network diagram. First, because fighters who had the same reference, facilitator, and arrived at the same place on the same day are most likely to know each other. But, second, because the number of fighters who are connected across all three networks serves as a proxy indicator for the amount of freedom fighters to self-organize and mobilize. Fighters who join ISIS together in large numbers is another signal, albeit imperfect, that they had greater freedom to mobilize. All other things being equal, for a high-risk activity, it is easier and more efficient to join in groups.[21] However, as we will see, other considerations might prevent these networks from growing too large in size.

Third, I will determine which characteristics of ISIS foreign fighters are most predictive of whether they share a tie with each other. To do this, I will use exponential random graph modeling (ERGM), which is a statistical model that analyzes network structure and allows us to make certain inferences about the attributes that predict whether two nodes share a tie.[22] The main focus of this study is the extent to which a fighters' geographic origin predicts his likelihood of sharing a tie with another fighter. How likely is a fighter from Europe to be connected to other fighters if they are from the same country? This ERGM analysis will also include other attributes, such as age, marital status, level of religious knowledge, foreign travel, previous experience as a Salafi Jihadist, professional history and educational background. These attributes are mainly "control" variables, although their presence also helps us assess the level of homophily in each network.

---

[20] Mathieu Bastian, Sebastian Heymann, and Mathieu Jacomy, "Gephi: an open source software for exploring and manipulating networks." (paper presented at the 3rd annual International AAAI Conference on Weblogs and Social Media, San Jose, California, USA, April, 2009).
[21] Walter Enders and Xuejuan Su, "Rational Terrorists and Optimal Network Structure" *Journal of Conflict Resolution* 51 no. 1 (2007): 33-57.
[22] Dean Lusher, Johan Koskinen, and Garry Robins, eds., *Exponential Random Graph Models for Social Networks* (Cambridge UK: Cambridge University Press, 2012); D.R. Hunter, M.S. Handcock, C.T. Butts, S.M. Goodreau, and M. Morris, "ergm: A Package to Fit, Simulate and Diagnose Exponential-Family Models for Networks." *Journal of Statistical Software* 24, no. 3 (2008): 1-29.

## III. Comparing Networks of ISIS Fighters from Europe and the Arabian Peninsula

This section analyzes the interconnectedness of fighters in the Arabian Peninsula and Europe samples in two parts. First, it will compare the statistics of connectedness of each network at different levels of degree restriction (denoted as 'L'). This is to determine whether clusters of fighters from the two regions behave differently as they get larger. Second, it will compare the strength of fighter ties in each network by identifying the fighters who share a tie across all three networks. Overall, the diagrams in Figure 1 and the statistics in Table 3 show that the Arabian Peninsula network is more interconnected than the fighters from Europe.

Table 3 – Comparison of Descriptive Statistics in Europe and Arabian Peninsula Networks

| | **Reference Network** | | **Facilitator Network** | | **Entry Network** | |
|---|---|---|---|---|---|---|
| | Europe | Arab Penin | Europe | Arab Penin | Europe | Arab Penin |
| **All Nodes:** | | | | | | |
| Av. Weighted Degree | 1.9 | 3.5 | 46.7 | 156.3 | 2.2 | 2.3 |
| Connected Comp. | 344 | 576 | 260 | 711 | 290 | 543 |
| Percent Isolates (Deg=0) | 53% | 40% | 47% | 22% | 36% | 30% |
| **Degree > 0** | | | | | | |
| Av. Weighted Degree | 3.9 | 5.8 | 87.8 | 200.8 | 3.4 | 3.2 |
| Connected Comp. | 69 | 154 | 17 | 260 | 104 | 226 |
| Percent of Nodes (Deg >0) | 47% | 60% | 53% | 78% | 64% | 70% |
| **Degree > 5** | | | | | | |
| Av. Weighted Degree | 7.7 | 10.9 | 98.1 | 206 | 7.4 | 6.8 |
| Connected Comp. | 10 | 24 | 8 | 241 | 11 | 25 |
| Percent of Nodes (Deg >4) | 16% | 25% | 47% | 76% | 16% | 16% |

Figure 1 – Comparison of Network Diagrams – Europe and Arabian Peninsula

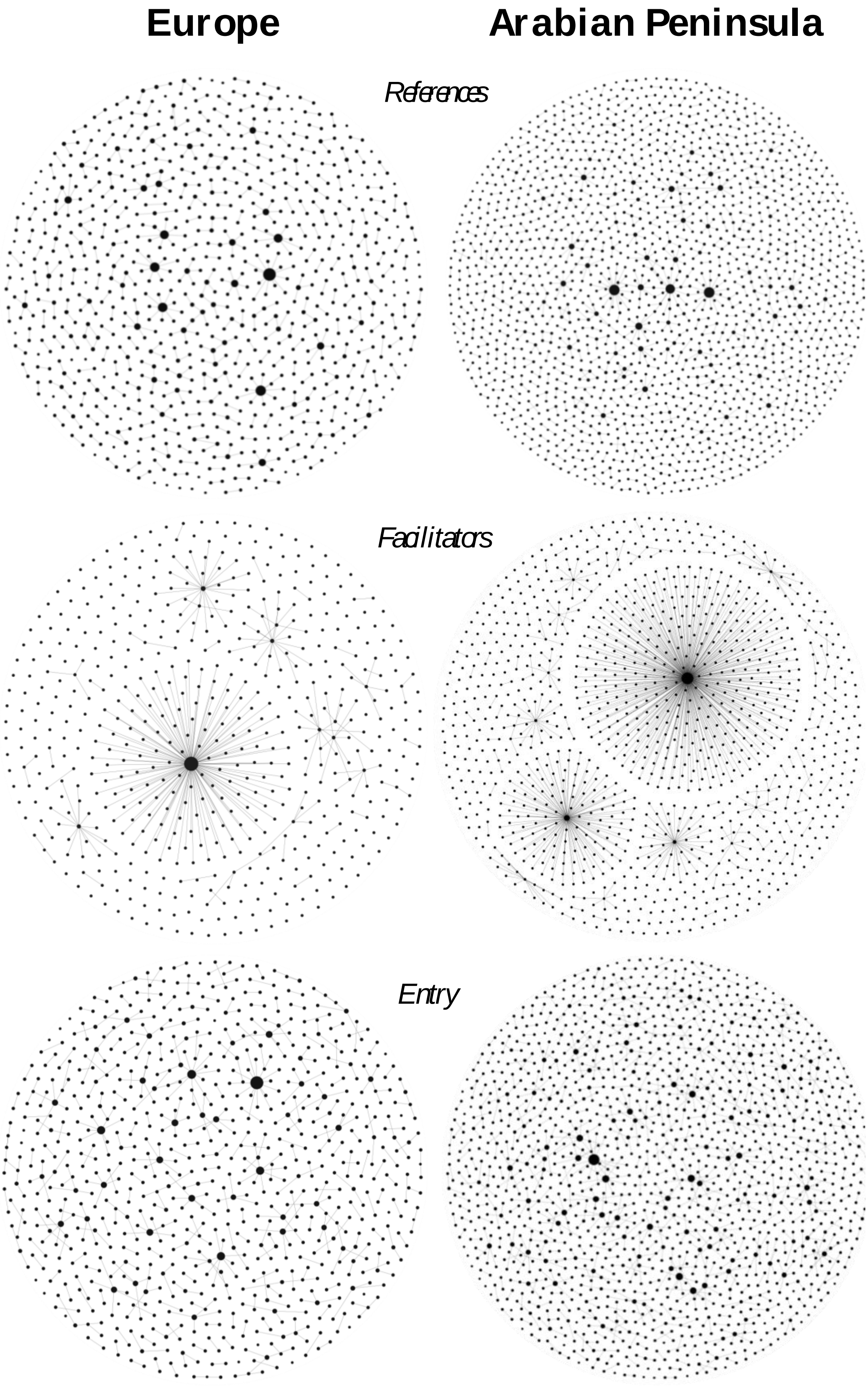

Table 3 shows that, based on descriptive network statistics, the network of Arabian Peninsula fighters is larger and more interconnected than those from Europe. There are three key statistics that describe the difference in the connectedness of the networks in the two regions: the average weighted degree, the number of connected components, and the percentage of isolates or nodes with a degree greater than or equal to a certain value. The average weighted degree is calculated by adding the number of edges for each node with consideration for their weight. For example, if a node has two edges, one weighted 1 and the other weighted 2, then it has a weighted degree of 3. That is (1+0) for the first edge, plus (1+1) for the second edge.[23] The number of connected components is the number of groups of interconnected nodes in a graph that include two or more connected nodes. The connected component statistic in Table 3 is not adjusted for the fact that the Arabian Peninsula has more than twice as many nodes (1059) as the Europe sample (519). Therefore, one would expect that the Arabian Peninsula sample would have approximately twice as many connected components at each level of degree restriction. The percentage of isolates, or the percentage of nodes with a degree of 0, shows the distribution of node connectivity.

When including isolates, fighters from the Arabian Peninsula have nearly two times the average weighted degree of the European fighters in the reference network. In the facilitator network, the nodes in the Arabian Peninsula facilitator network have an almost 3.5 times larger weighted degree average than the European fighter sample, but the two regional networks converge in terms of their connectivity when considering their entry networks. The explanation for this convergence will be treated later in a sub-section of section IV titled "ISIS Managing Europe Fighter Arrivals." It is also possible that the convergence for entry networks is because there is a limited number of people who can be transported on any given day through the smuggling routes used to cross into Syria from Turkey.[24]

[23] Borgatti et al (see note 27 above)

The number of connected components is larger for the Arabian Peninsula in all three graphs than that of Europe. But this is not always more than double, which means that they have proportionately roughly the same number of connected components. In each of the networks, the Arabian Peninsula sample has fewer isolates than the European sample; however, these percentages converge in the entry network.

By restricting the size of the network by degree (L) – first to nodes with an L greater than 0 (i.e., nodes that have at least one tie with another) and then to nodes with a degree of 5 or greater (i.e., nodes that have at least five ties) – the differences between the Arabian Peninsula sample and the Europe sample remain relatively consistent. In the reference and facilitator networks, the Arabian Peninsula sample is more interconnected: a larger percentage of its nodes have degrees greater than 0 and 5 than those in Europe, and each one of those nodes generally has more ties, as evidenced by the larger average weighted degree. For the connected components, when $L>0$ and $L>4$, the Arabian Peninsula sample has more than twice as many connected components, and, in the facilitator network, substantially more than twice the number of connected components. This indicates that the nodes at higher degree levels are not just more connected, but that there are more groupings of connections. However, the difference in network interconnectedness between the regions in the entry networks remains negligible, even as the degree level increases.

---

[24] Tim Arango and Eric Schmitt, "A Path to ISIS, Through a Porous Turkish Border," *New York Times*, March 9, 2015, https://www.nytimes.com/2015/03/10/world/europe/despite-crackdown-path-to-join-isis-often-winds-through-porous-turkish-border.html (accessed January 5, 2019).

### *"Strong" Inferences - How Many Fighters Shared the Same Referral, Facilitator, and Arrived at the Same Place on the Same Day?*

This section will look at the number of fighters from Europe and the Arabian Peninsula who were most likely to know each other – those who shared the same referral, facilitator, and who arrived on the same day at the same time. These results help us get a better sense for the level of freedom people had to self-organize and mobilize in each region. The more people shared a referral, facilitator, and arrival date and location, the more likely they were to know each other and the easier it would be for them to travel together. For the purposes of brevity, I call these ties "strong" ties between fighters.

Figure 2– Strong Tie Networks – Europe and Arabian Peninsula

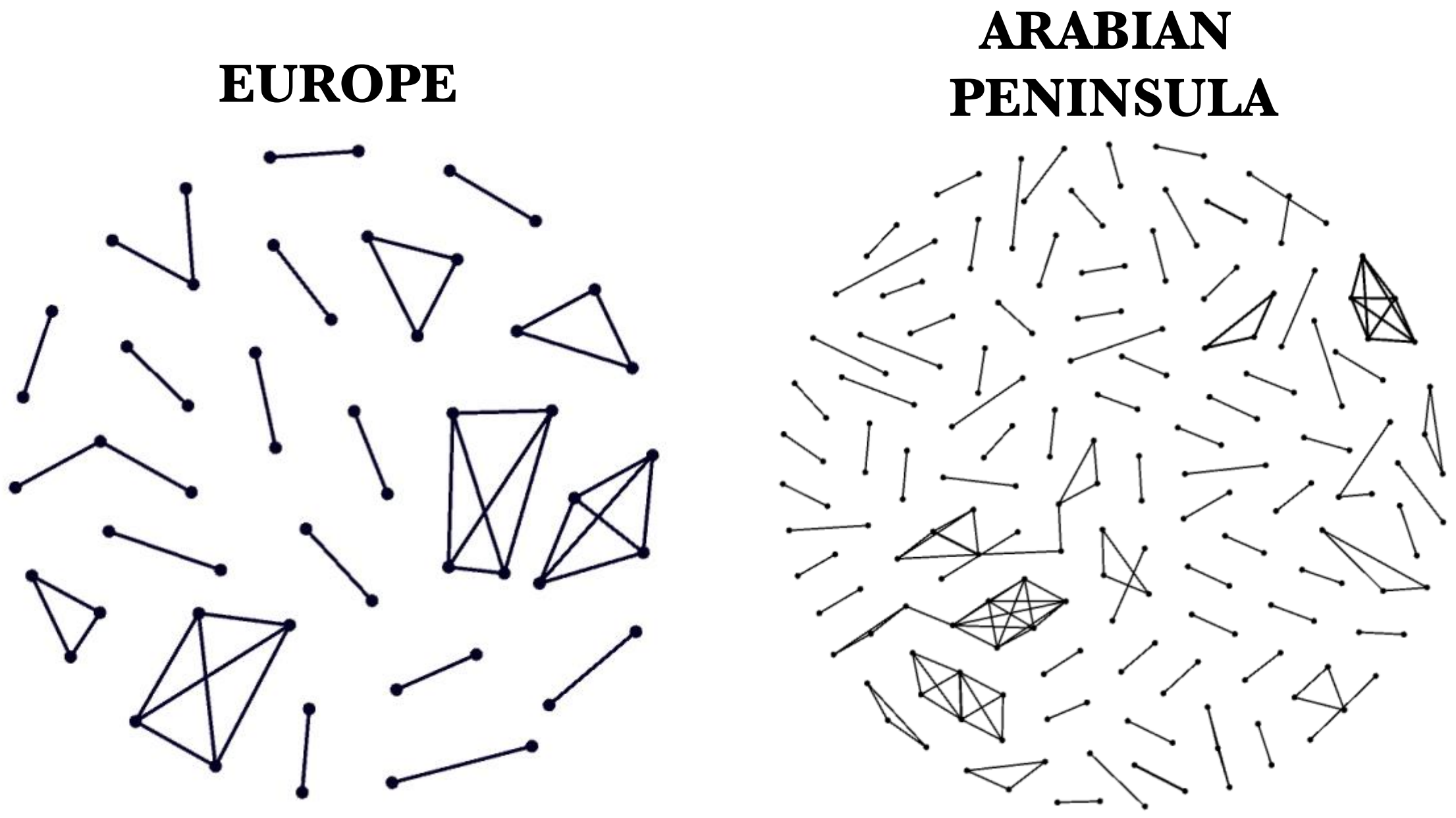

Table 4 – Comparison of Strong-Tie Networks in Europe and Arabian Peninsula

| Descriptive Statistic | Arabian Peninsula | Europe |
|---|---|---|
| Average Weighted Degree | 5.0 | 5.0 |
| Connected Components | 85 | 21 |
| Percent of Nodes w/ an Edge ≥ 3 | 19% | 10% |

To analyze the strength of ties between fighters, I combined all three networks into one network and weighted the edges 1-3 depending on how many answers each fighter with a tie shared in common.[25] I then restricted the networks to only show fighters who share a tie with an edge weight of at least three, and these are shown in Figure 2.

As shown in Table 4, the Arabian Peninsula has more than double the proportion of fighters who have strong-tie relationships than those from Europe. It also has four times the number of connected components, and a comparable average weighted degree. This suggests that fighters from the Arabian Peninsula are not only more connected than those from Europe, but also more "strongly" connected. This has analytical value: that the Arabian Peninsula network has more "strong" ties means it was more likely fighters from there were able to organize openly and collectively. They could remain connected by their referral and also travel together to Syria via the same facilitator and arrive on the same day at the same place.

**IV. What Predicts Whether Two Foreign Fighters "Know" Each Other?**

We now know that Arabian Peninsula foreign fighter networks were larger and more freely forming. But what factors predict how they form? This section will compare ERGM results for fighter networks on two levels: first, it will test country-level origins of fighters from Europe and the Arabian Peninsula. Second, it will test how strongly province-level origins predict ties among fighters from France and Saudi Arabia. This is to determine how the geographic dispersion of fighters varies between countries in Europe and the Arabian Peninsula.

ERGMs are like regressions for network data. They help identify the attributes that are most predictive of a tie between fighters. To describe these correlations, I use the Lusher et al method for deriving conditional odds from ERGM coefficients.[26]

$$(\Pr(X = x|Y = y) = (\tfrac{1}{k}) \exp(\theta^T z(x) + \theta_a \textstyle\sum_{i<j} x_{ij}(y_i + y_j) + \theta_b \sum_{i<j} x_{ij} y_i y_j)$$

[25] Some fighters have an edge weight greater than three if they 1) share a tie in all three networks and 2) share multiple answers in common to the same question (e.g., multiple references).
[26] Lusher et al (see note 31 above).

Where the conditional odds of the tie between nodes i and j is equal to the exponentiated sum of the coefficients θa and θb – the attributes shared by i and j - over the constant, while the $\exp(\theta^T z(x))$ component is the conditional odds of a tie when the two nodes do not share any attributes. The constant is depicted in an ERGM result as the "edges" variable (see Table 5 below).

## *ERGMs of European and Arabian Peninsula Networks*

In the European sample, the reference and entry networks show a very strong positive correlation with regards to the country of origin, while the facilitator network shows few attributes being predictive of shared ties between fighters (see Table 5 for regression results below). In the reference network, with all else being equal, fighters in the Europe sample were more than 11 times more likely to share a tie if they were from the same country. This connection is even stronger in the entry network, where fighters were 55 times more likely to share a tie if they were from the same country. For the reference and entry networks among European fighters, it appears that many attributes that might apply in normal conditions of homophily – birth year, educational background, and previous jihad experience – are also strongly indicative of shared ties. However, with the exception of religious knowledge, all of these attributes decline in relevance for the entry network. Virtually no attribute indicates a shared tie in the facilitator network of the Europe sample, with the modest exception of the country of residence.

Table 5 – Results of ERGMs for Arabian Peninsula and Europe Networks

| | Reference Network | | Facilitator Network | | Entry Network | |
|---|---|---|---|---|---|---|
| | Europe | Arab. Pen. | Europe | Arab. Pen. | Europe | Arab. Pen. |
| ***Country of Residence*** | ***2.41 (0.12)**** | ***-0.38 (0.1)**** | ***0.37 (0.03)**** | ***0.12 (0.01)**** | ***4.01 (0.13)**** | ***0.04 (0.1)*** |
| Birth Year | 0.82 (0.15)*** | 0.01 (0.1) | 0.03 (0.05) | -0.13 (0.02)*** | 0.15 (0.18) | 0.03 (0.12) |
| Nationality | 1.15 (0.11)*** | 0.38 (0.1)*** | -0.09 (0.04)* | -0.16 (0.01)*** | 0.29 (0.09)** | -0.07 (0.11) |
| Education | 0.66 (0.09)*** | 0.06 (0.05) | 0.42 (0.02) | 0.04 (0.01)*** | 0.15 (0.09)** | 0.17 (0.06)** |
| Religious Knowledge | 0.37 (0.01)*** | -0.13 (0.05)** | -0.3 (0.02) | 0.08 (0.01)*** | 0.48 (0.09)*** | 0.03 (0.06) |
| Profession | 0.13 (0.10) | -0.03 (0.06) | -0.13 (0.02) | -0.01 (0.01) | 0.08 (0.1) | 0.01 (0.07) |
| Prev. Jihad Experience | 0.77 (0.13)*** | 0.05 (0.06) | 0.04 (0.02)· | 0.36 (0.01)*** | -0.09 (0.1) | 0.05 (0.07) |
| Marital Status | 0.004 (0.1) | 0.22 (0.05)*** | -0.13 (0.02)*** | 0.06 (0.01)*** | 0.02 (0.09) | 0.06 (0.06) |
| Countries Visited | 0.66 (0.1)*** | 0.07 (0.06) | 0.02 (0.02) | 0.01 (0.01) | 0.41 (0.1) *** | 0.15 (0.07)* |
| *Edges* | *-8.02 (0.15)**** | *-5.85 (0.06)**** | *-2.29 (0.02)**** | *-2.09 (0.01)**** | *-7.76 (0.15)**** | *-6.3 (0.08)**** |
| AIC/BIC | 5019/5117 | 24457/24580 | 185557/185655 | 774746/774870 | 184400/184498 | 17043/17155 |

*Standard errors in parenthesis; '***' p < 0.001, '**' p < 0.01, '*' p< 0.05, '.' p <0.1; Europe (N=519); Arabian Peninsula (N=1059)*

Meanwhile, the network analysis of the fighters from the Arabian Peninsula shows that country of origin has almost no predictive power for tie formation. In the reference network, country of origin negatively predicts fighter ties (-0.38) and is only modestly positive in the facilitator and entry networks (0.12 and 0.04 respectively). This means that, all else being equal, a fighter from Saudi Arabia is just as likely to share a tie with another Saudi as he is with, for example, a fighter from Kuwait. Meanwhile, some personal attributes predict ties with statistical significance between fighters in more than one of the three networks, such as marital status and educational background.

***ISIS Managing Europe Fighter Arrivals***

The explanation for why the results appears different for the facilitator network among fighters from Europe is that there appears to be a deliberate sorting of fighters from Europe. This is likely done by ISIS recruiters, who appear to have arranged for fighters from certain countries to arrive at ISIS territory on a specific day and location. This would explain two results in the previous findings: first, that few attributes predict ties in Europe's facilitator network, and second, that the country of origin attribute so strongly predicts ties in Europe's entry network results (see Figure 3). Since few attributes predict ties in the facilitator network, connections in this network are likely organized by an external actor – in this case, it is likely to be the ISIS bureaucracy.

Figure 3 – ISIS-Organized Arrivals for European Fighters

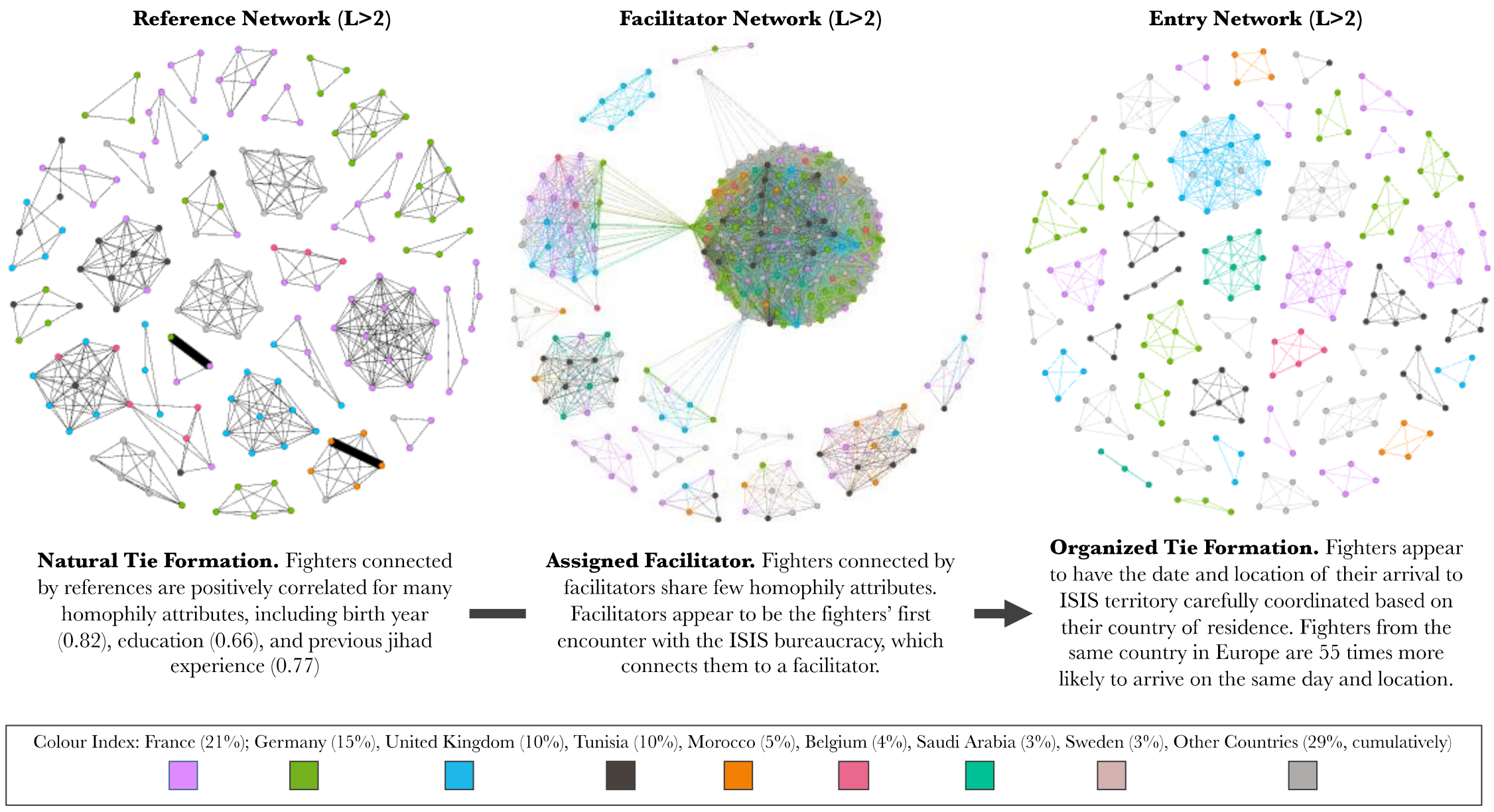

The facilitator is an ISIS operative who works out the logistics of foreign fighter entry into ISIS territory.[27] Reporting has described how ISIS sometimes manages the process of fighter arrivals, rather than letting the fighters plan the trip themselves.[28] Given this, it is likely that facilitators managed the process of European ISIS joiners arriving in Syria and organized them by country of origin.

As a result of this finding, I treat the facilitator and date/location of arrival networks as being at least partially managed by ISIS officials in Syria. The reference network, meanwhile, which shows the most amount of homophily, is likely still a network formed by fighters without the influence of an external actor (e.g., ISIS officials). These results are important to mention, but do not change the fundamental differences between fighter networks from Europe and the Arabian Peninsula.

[27] Dodwell et al (see note 25 above).
[28] Alessandria Masi and Hanna Sender, "How Foreign Fighters Joining ISIS Travel to the Islamic State Group's 'Caliphate'," *IB Times*, March 3, 2015, http://www.ibtimes.com/how-foreign-fighters-joining-isis-travel-islamic-state-groups-caliphate-1833812 (accessed December 19, 2018).

## V. Mechanisms: What Might Explain These Results?

The key finding of this research is that there are structural differences in the composition of the social networks of foreign fighters. Fighters from the Arabian Peninsula had more ties, more strong ties, and their ties were more geographically dispersed than those from Europe. Yet despite these differences, all fighters studied here joined the same organization in the same conflict at roughly the same time. It shows that the recruitment process for new joiners varied significantly within a single organization.

Fighters are more interconnected in the Arabian Peninsula than they are in Europe in terms of the number of ties, the strength of their ties, and the number of clusters of large ties. In Europe, the networks of recruits are more disparate and disconnected, suggesting that recruitment is more likely to occur on a one-to-one basis than in the Arabian Peninsula, where interconnected networks suggest a more social and interpersonal recruitment process.

There are several possible mechanisms that could explain the difference in networks between fighters from Europe and the Arabian Peninsula. One may be that the Muslim majority societies of the Arabian Peninsula simply offer greater opportunities for potential recruits to meet each other: there are more mosques and preachers, and a larger proportion of state and private resources spent on religious activity.

But ethnographic research into European neighborhoods such as Molenbeek, one of the largest Muslim communities in Brussels, might challenge this explanation. Fraihi, for example, found high concentrations of potential recruits for radical Salafist movements among local youth almost a decade before ISIS was formed.[29] Her finding suggests that while the absolute number of resources potential recruits could use to meet is greater in the Arabian Peninsula than in Europe, it may be that the concentration of such resources (i.e., in neighborhoods like Molenbeek) is actually greater. As such, this explanation is likely incomplete.

A second reason for the differences in connectivity between fighters from Europe or the Arabian Peninsula may be shared language. Although most residents of Europe can travel freely through its Schengen Area, residents of its different countries speak different languages. This might constitute a barrier to the opportunities that, for example, a person living in France might face to meeting someone from Germany. These same barriers do not apply in the Arabian Peninsula, where the shared language is Arabic.

On closer examination of the results, there are three reasons why I believe language and accessibility explanations are incomplete. The first is that the results of fighters from Europe show that the do not share transnational ties with fighters who might share their language abroad (i.e., Moroccan residents in Belgium are unlikely to share a tie with Moroccans from Morocco). If language determined tie formation between ISIS foreign fighters, we would expect there to be many more ties between Moroccan fighters living in Europe and Moroccan fighters who travel to Europe (these are recorded as "Europe" in the data – see methods section). But the results do not show a large number of these types of transnational ties.

[29] Fraihi (see note 5 above)

Figure 4 – Hierarchy in Arabian Peninsula Fighters, Decentralized Networks in Europe

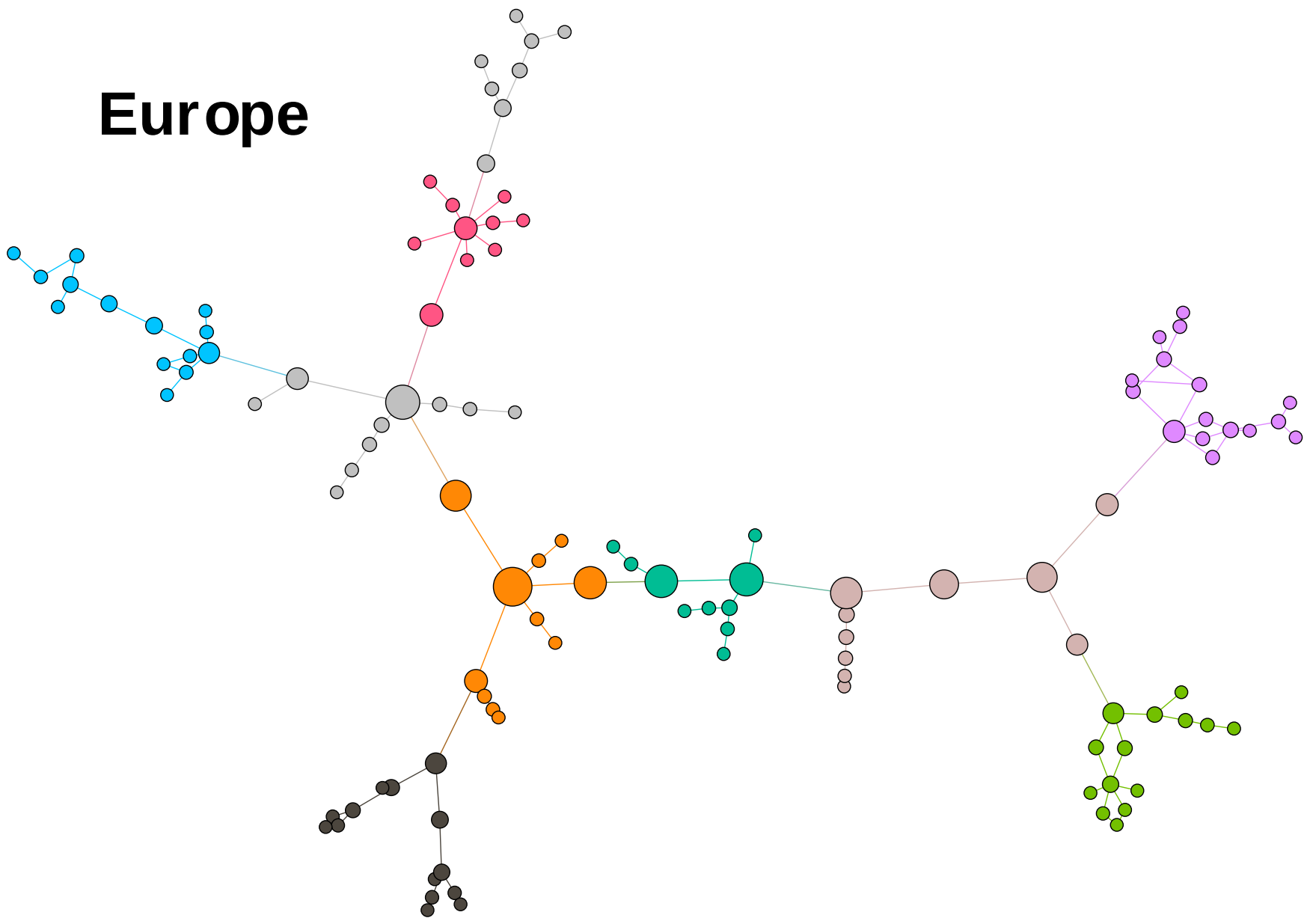

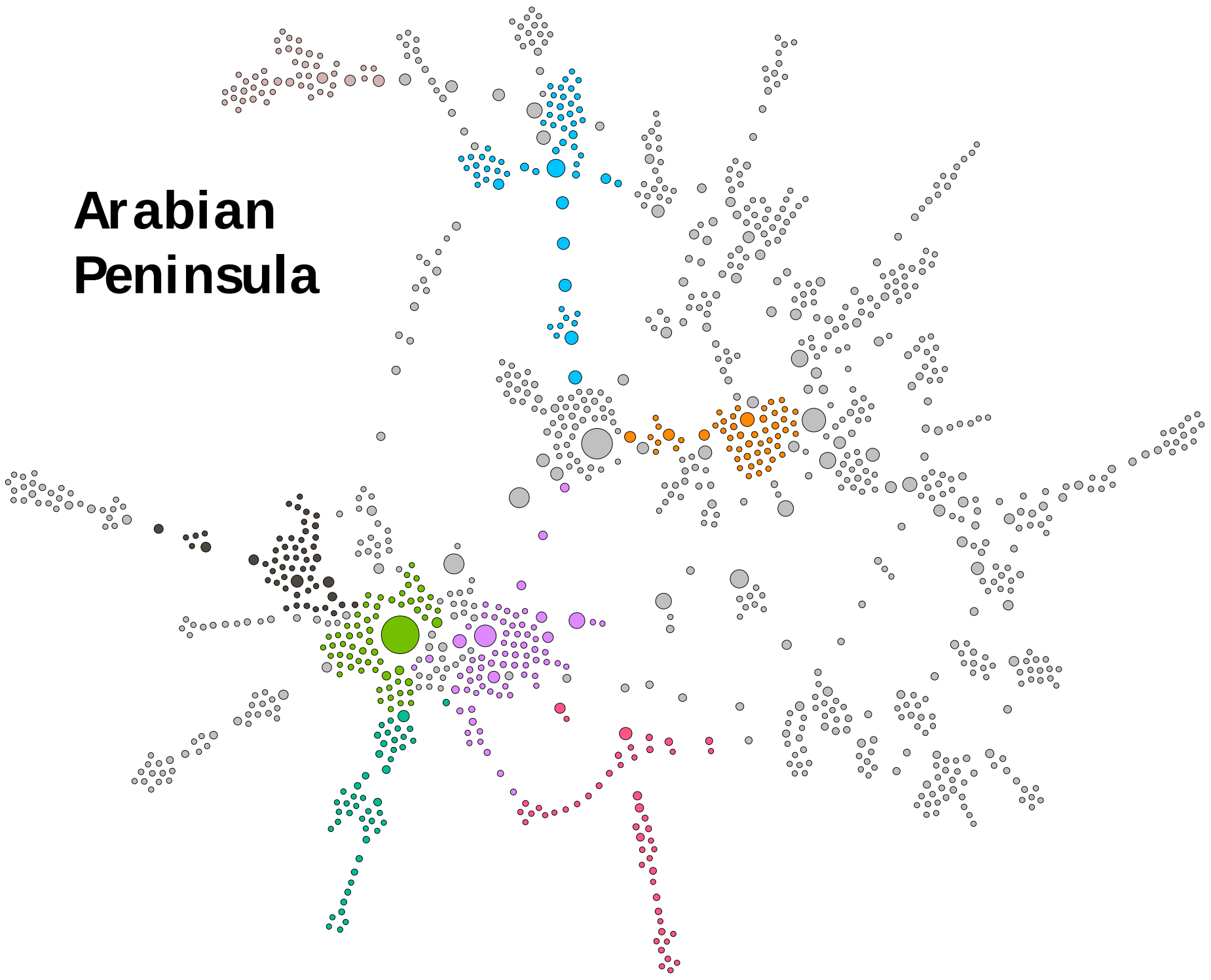

The second reason relates to network hierarchy. Network hierarchy suggests a freedom to organize: the large nodes in the Arabian Peninsula diagram are brokers between large communities of fighters. If linguistic commonalities were the main reason driving the differences between ISIS foreign fighter networks in Europe and the Arabian Peninsula, the hierarchy in the Arabian Peninsula network would be less pronounced (shown in Figure 4). The network diagrams show the structure of the main component of the entire network of fighters from Europe and the Arabian Peninsula. Whereas the Arabian Peninsula resembles a network where there is some clear hierarchy among certain nodes which are far more central to the model (sometimes called a "scale-free" network). The central nodes in Figure 4 are highlighted by their size – larger nodes possess greater "betweenness centrality" (they are critical nodes one would need to travel through if one wanted to move across the network). That this hierarchy so clearly exists in the Arabian Peninsula network, as opposed to that from Europe, underscores the fact that it was not just shared language in the Arabian Peninsula but, likely, a greater freedom to mobilize that contributed to how the network of foreign fighters formed. Linguistic commonalities might account for the fact that there are a large number of ties between fighters, but it cannot explain the hierarchical structure that appears among fighters who join from the Arabian Peninsula.[30]

The third reason why shared language is unlikely to explain why European foreign fighters are more likely to be connected to each other by country than those from the Arabian Peninsula is that these same findings hold up when we zoom into specific countries. Appendix VI shows how fighters from France are much more likely to be connected to each other based on where they come from inside France than those fighters from Saudi Arabia from within Saudi Arabia. This finding serves as a "robustness" check that confirms the trends we found in sections III and IV.

These three factors: 1) the hierarchy of networks in the Arabian Peninsula, 2) the lack of a transnational dimension in European networks, and 3) the fact that these trends also exist inside European countries – demonstrate that language and accessibility do not explain why fighters in the Arabian Peninsula were freer to organize and mobilize. This inference would confirm previous research on the secrecy-efficiency tradeoff of terrorist groups. Secret groups need hub and spoke models of organizing, with decentralized networks that avoid detection or disruption but cannot grow. More efficient organizational models, like scale-free, hierarchical networks, only form among members of an insurgent or covert group when they are freer to organize.[31]

[30] Note: the network diagrams in Figure 4 combined the reference and date/location of entry networks. As demonstrated in Figure 3, the facilitator network does not reflect fighters' self-organizing is therefore excluded.

## VI. Conclusions, Limitations, and Policy Recommendations

This paper compares the network of fighters who joined the Islamic State of Iraq and Syria (ISIS) in 2013-14 from Europe and the Arabian Peninsula in order to test whether there are structural differences in their recruitment. It is the first comparative network analysis of foreign fighters from different regions that joined the same group in the same conflict at the same period of time. This study found five important results that are relevant for future research and policy.

First, the network of fighters from the Arabian Peninsula was larger, more hierarchical, and more geographically dispersed than those from Europe. Whereas the network of European foreign fighters was smaller, more geographically constrained inside their country of origin, and more decentralized in terms of their network structure. In addition, ISIS may have deliberately managed European foreign fighters to arrive by their country of origin, another factor highlighting their greater concern for security.

Second, the most likely explanation for these network differences are due to the different contexts in which recruitment occurred. The law enforcement scrutiny in Europe meant that networks of fighters were smaller, more local, and more decentralized. This localism was a way to maintain the secrecy of the fighters' identities and intentions. Meanwhile, networks were larger, geographically dispersed, and hierarchical in the Arabian Peninsula, where recruitment occurred more freely. Alternative explanations, such as cultural or linguistic differences, are considered and rejected in the paper.

[31] Enders and Su (see note 32 above); Frederic S. Pearson, Isil Akbulut & Marie Olson Lounsbery (2017) Group Structure and Intergroup Relations in Global Terror Networks: Further Explorations, Terrorism and Political Violence, 29:3, 550-572, DOI: 10.1080/09546553.2015.1058788

Third, the ISIS foreign fighter recruitment process varied significantly. It would be a mistake to treat ISIS as a homogenous group whose fighters share similar recruitment pathways. Researchers who study recruitment and radicalization should carefully frame their findings in terms of the social, political, and legal contexts in which their research takes place.

Fourth, European recruits were more isolated than those from the Arabian Peninsula, but it is unlikely they joined ISIS on their own. Yet one-on-one recruitment at such a large scale is unlikely to be carried out in person and more likely mediated online. This theory is worth testing in the future: whether the smaller, weaker fighter networks from Europeans meant that they were more likely recruited online than those in the Arabian Peninsula.

Fifth, the extent of the networks shows that offline/online recruitment likely plays a greater role on smaller networks with a larger proportion of isolated nodes. In Europe, Reynolds and Hafez found less than 5% of German fighters in their study were confirmed to have been radicalized online.[32] But Kennedy and Weimann emphasize the myriad uses insurgents might have to recruit on the Internet. As the authors write, their goal is "not so much to assess the merits of the Internet as a replacement for physical interaction, but rather to demonstrate how terrorists utilize the Internet" to build connections to potential recruits.[33] In light of this argument, it might be more useful to think about the two influences – in person and online – as complements rather than substitutes. Some fighters are exclusively recruited in person, while others might be exclusively recruited online. More often, it is a combination of the two working in tandem. The fact that Europe's fighters appear more disconnected in comparison to those from the Arabian Peninsula suggests that the recruitment process there was more frequently one-on-one. And given the scale and speed of ISIS recruitment, it is likely that much more of this one-on-one recruitment was mediated through the Internet. Those isolates from Europe may not have exclusively used to Internet for their mobilization, but they may have been more likely to rely on the Internet than the ones who were more interconnected to other joiners.

---

[32] Reynolds and Hafez (see note 20 above)
[33] Jonathan Kennedy and Gabriel Weimann, "The Strength of Weak Terrorist Ties," *Terrorism and Political Violence* 23 no. 2 (2011): 201-212, DOI: 10.1080/09546553.2010.521087, p. 207.

### *Limitations of the Study*

This study has three key limitations. First, the study relies on self-reported personal information to a terrorist organization. The foreign fighters analyzed in this study were less likely to lie than if they were interviewed by a journalist or, possibly, law enforcement, but there is likely some information that a recruit withheld purposively (e.g., due to secrecy) or accidentally (e.g., due to language barriers). In generating this network data, a subject that offers the name of three references is more likely to share a tie with other fighters than one who submits no name at all. Therefore, it is entirely possible that some of the most interconnected nodes in the network diagrams are simply the people who provided the most detailed information in their form. Important connections in the network may not have been captured because of incomplete or inaccurate responses.

Second, the two-mode networks analyzed in this study likely have varied degrees of accuracy in terms of their predicting actual relationships between fighters. Unlike the detailed report on German foreign fighters by Reynolds and Hafez, this paper does not have details on how these ties formed.[34] We do not know, for example, how many ties formed based on shared interest in joining ISIS or whether they were pre-existing friendships or familial ties. Two-mode networks infer ties, and for reference networks, the relative homophily of the attributes inferring ties between fighters, and information about the reference himself, suggests that this network is the most predictive of an actual shared tie between two fighters in the study. That many references are likely local to the context from which the fighter originates means that they are likely to be personally known in some capacity by both fighters. The properties of triadic closure that underpin the assumptions of two-mode networks – that if node A and node B both know node C very well, then they are eventually likely to know each other as well – are strong in the reference network case.[35] This is hardly evident in the facilitator network, and only slightly evident in the arrival network.

Third, one of the most significant challenges in carrying out this research was reconciling the myriad inconsistencies with which answers were recorded in the forms. These inconsistencies were mostly minor and required careful hand coding to ensure accuracy. But inconsistencies were sometimes more complicated, such as the transcription of people's names. Some recruits reported their facilitator or reference by his real name, others by his *kunia* (nom-de-guerre). And some well-known fighters have many different kunias, such as Omar al-Shishani, a senior ISIS commander who was known by at least 13 different names.[36] I reconciled these different reference names for some people in the data, but likely not all. Therefore, it is possible that the networks involving shared reference to an individual (e.g., the facilitator and reference networks) underestimate the total number of links between fighters, although this effect is unlikely to be biased against one country or region.

---

[34] Reynolds and Hafez (see note 20 above)
[35] Borgatti et al (see note 27 above)
[36] Author Unknown, "Wanted: Tarkhan Tayumurazovich Batirashvili," *U.S. Department of State – Rewards for Justice,* 2016, https://web.archive.org/web/20160311070758/https://www.rewardsforjustice.net/english/tarkhan_batirashvili.html (accessed January 4, 2019).

### ***Policy Recommendations***

Two policy recommendations flow from these findings – one clear and the other more subtle. The clear finding of this research is that there are structural differences in the composition of social networks of people who join terrorist organizations from different regions. This is demonstrated in the fact that fighters from the Arabian Peninsula are different than fighters from Europe in the breadth, depth, and dispersion of their ties, even though they joined the same organization at roughly the same time. The implications of this are profound: that even within a single organization, the recruitment process for new joiners varies significantly. It would therefore be a mistake to treat ISIS as a homogenous group whose fighters share similar recruitment pathways. Research and policy on focused on the recruitment and radicalization of fighters to ISIS in particular, and transnational terrorist groups in general, should carefully frame their approach in terms of the social, political, and legal contexts in which their research or policy takes place. Given the fundamentally different recruitment processes experienced by fighters from Europe and the Arabian Peninsula, contextualizing their origins is essential.

The second finding is more subtle: just as terrorist groups face a secrecy-efficiency tradeoff in their operations and in their recruitment, counterterrorism efforts also face efficiency tradeoffs in fighting the group.

Figure 5 – Notional Hierarchical and Decentralized Networks

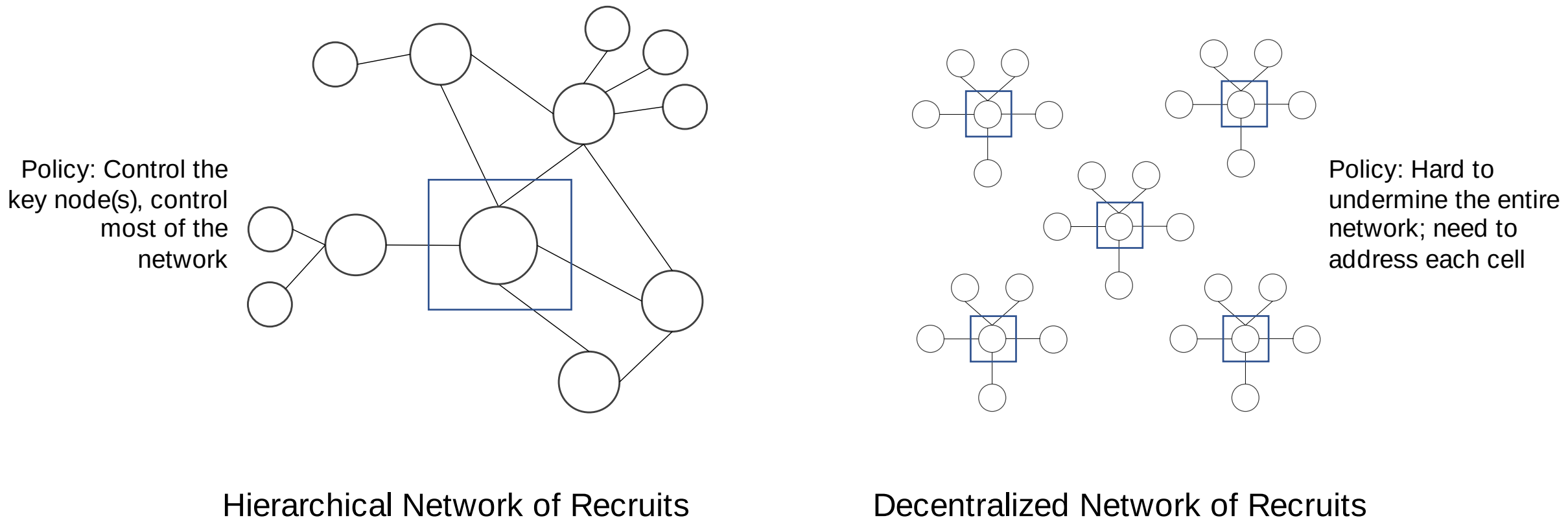

The group with the larger, more hierarchical recruitment structure shown in Figure 5 is easier to control. Control the nodes in the network with the most influence, and one can control the entire network. But the risk with these types of networks is they pose a larger, more systemic threat and are likely to emerge in places with weaker counterterrorism infrastructure. In such places, the policy is generally to find ways to control the key nodes in the network. This is the most efficient way of addressing the threat with limited counterterrorism resources, because it involves the least amount of effort (i.e., finding ways of dissuading key individuals from threatening the state) for the desired outcome (i.e., minimizing the terrorist threat). The problem with this approach is that losing control of these nodes creates a significant threat. A good real-world example of this type of model is the network of ISIS supporters in Mosul in 2014: in July, seemingly overnight, these networks were activated, and they overwhelmed Iraqi security forces *en route* to ISIS capturing the city.[37]

Instead, bilateral and multilateral efforts should be made to help states that face system-wide risks of terrorist groups push these groups to operate in secret. In doing so, they force these groups to recruit on the secrecy side of the secrecy-efficiency spectrum, which means small cells recruiting fewer followers and only in certain key locations. This is a good outcome because it reduces the potential size of a terrorist group's followers. But if this transformation, which comes at a high cost, indeed occurs, it creates a new risk. Instead of threatening the state's control of a certain city or region, decentralized terrorist group cells that are resilient to disruption pose the threat of one-off attacks that can still have a significant impact (i.e., ISIS attacks from 2015-20 that occurred in European cities like Paris, London, Berlin, and Barcelona). Efforts to help states transform their threat from an efficient and hierarchical one to a decentralized and resilient threat of one-off attacks must transform their counterterrorism efforts to identify this transformation and respond to the evolving threat.

[37] Yasir Abbas and Dan Trombly, "Inside the Collapse of the Iraqi Army's 2nd Division," *War on the Rocks,* July 1, 2014.

## Appendix I – Question Form

الإدارة العامة للحدود

بسم الله الرحمن الرحيم

الدولة الإسلامية في العراق والشام

الإدارة العامة للحدود

بيانات مجاهد

| | | |
|---|---|---|
| 1 | الإسم واللقب | |
| 2 | الكنية | |
| 3 | إسم الأم | |
| 4 | فصيلة الدم | |
| 5 | تاريخ الولادة و الجنسية | |
| 6 | الحالة الإجتماعية | أعزب (1) متزوج ( ) عدد الأطفال ( ) |
| 7 | العنوان و مكان الإقامة | |
| 8 | التحصيل الدراسي | |
| 9 | المستوى الشرعي | طالب علم ( ) متوسط ( ) بسيط (1) |
| 10 | ماهي مهنتك قبل المجيء ؟ | |
| 11 | البلدان التي سافرت إليها وكم لبثت بها؟ | |
| 12 | المنفذ الذي دخلت منه ؟ والواسطة ؟ | تل ابيض |
| 13 | هل لديك تزكية ومن من ؟ | |
| 14 | تاريخ الدخول ؟ | |
| 15 | هل سبق لك الجهاد ؟ وأين ؟ | |
| 16 | مقاتل أم إستشهادي أم إنغماسي ؟ | |
| 17 | الإختصاص ؟ | مقاتل ( ) شرعي ( ) أمني ( ) إداري ( ) |
| 18 | مكان العمل الحالي | |
| 19 | الأمانات التي تركتها ؟ | جواز سفر |
| 20 | مستوى السمع والطاعة ؟ | |
| 21 | العنوان الذي نتواصل معه ؟ | |
| 22 | تاريخ القتل و المكان | |
| 23 | ملاحظات الانتماءات والحياة العامة والاعتقالات ومطلوب لا | |

الإدارة العامة الحدود

الدولة الإسلامية في العراق والشام _ سري _

**Appendix II – Question Form Translated**

Each form contained the following fields:

1. Full name
2. Kunya [Nickname / Nom de guerre]
3. Mother's name
4. Blood type
*5. Date of birth and citizenship
*6. Marital status: [check boxes for] Single, Married, Number of children
*7. Address and place of residence (NOTE: this was geo-referenced by country and province)
*8. Education level
*9. Level of sharia expertise: [check boxes for] Advanced Student, Intermediate, Basic
*10. Occupation prior to your arrival
*11. Countries visited and time spent in each
**12. Point of entry and facilitator
**13. Do you have a recommendation, and from whom?
**14. Date of entry
*15. Have you engaged in jihad before, and where?
16. [Do you want to be a] fighter, istishhadi [suicide bomber], or inghimasi [suicide fighter]?
17. Specialty: [check boxes for] Fighter, Sharia [official], Security [personnel], Administrative
18. Current work location
19. Personal belongings that you deposited
20. Level of understanding [of orders] and obedience
21. Address where [point of contact] can be reached
22. Date and location of death
23. Notes

* - Considered as a node attribute in this paper

** - Considered as an edge definition in this paper (i.e., two fighters who had the same answer to this question shared a tie)

# Appendix III – Fighters from Europe

| Europe-Wide Sample (note - non-European countries include those with one at least one fighter reporting self-report travel to a European country) | | | France Sample (note - all fighters reported residing in France) | | |
|---|---|---|---|---|---|
| **Country of Residence** | Count | Percent of Total | Province of Residence | Count | Percent of Total |
| **Albania** | 7 | 1.3% | Auvergne-Rhône-Alpes | 9 | 8.3% |
| **Algeria** | 6 | 1.2% | Île-de-France | 30 | 27.8% |
| **Australia** | 2 | 0.4% | Provence-Alpes-Côte d'Azur | 10 | 9.3% |
| **Belgium** | 22 | 4.2% | Hauts-de-France | 5 | 4.6% |
| **Canada** | 5 | 1.0% | Languedoc-Roussillon | 2 | 1.9% |
| **Denmark** | 12 | 2.3% | Bourgogne-Franche-Comté | 3 | 2.8% |
| **Egypt** | 7 | 1.3% | Occitanie | 13 | 12.0% |
| **Finland** | 5 | 1.0% | Listed location unknown | 6 | 5.6% |
| **France** | 108 | 20.8% | Normandy | 2 | 1.9% |
| **Germany** | 79 | 15.2% | Not listed | 10 | 9.3% |
| **Holland** | 10 | 1.9% | Nord-Pas de Calais | 1 | 0.9% |
| **Iraq** | 2 | 0.4% | Centre-Val de Loire | 2 | 1.9% |
| **Kosovo** | 12 | 2.3% | Grand Est | 14 | 13.0% |
| **Kuwait** | 9 | 1.7% | | | |
| **Libya** | 5 | 1.0% | | | |
| **Macedonia** | 7 | 1.3% | | | |
| **Morocco** | 27 | 5.2% | **Europe-Wide Sample (cont. from left)** | | |
| **Norway** | 3 | 0.6% | Afghanistan, Azerbaijan, Bahrain, Bosnia, Iceland, Italy, Lebanon, Martinique, Netherlands, Palestine, Romania Switzerland, Syria, Trinidad and Tobago, Yemen | All had 1 fighter | 3.0% (sum of all 15 countries) |
| **Poland** | 2 | 0.4% | | | |
| **Qatar** | 2 | 0.4% | | | |
| **Russia** | 10 | 1.9% | | | |
| **Saudi Arabia** | 17 | 3.3% | | | |
| **Spain** | 10 | 1.9% | | | |
| **Sweden** | 13 | 2.5% | | | |
| **Tanzania** | 2 | 0.4% | | | |
| **Tunisia** | 51 | 9.8% | | | |
| **Turkey** | 10 | 1.9% | | | |
| **United Kingdom** | 53 | 10.2% | | | |
| **USA** | 3 | 0.6% | | | |

# Appendix IV – Fighters from Arabian Peninsula

| Arabian Peninsula-Wide Sample (note = non-AP countries include those who self-report travel to an AP country) | | | Saudi Arabia Sample (note = all fighters included self-report residing in Saudi Arabia) | | |
|---|---|---|---|---|---|
| **Country of Residence** | **Count** | **Percent of Total** | **Province of Residence** | **Count** | **Percent of Total** |
| **Algeria** | 4 | 0.4% | **Mecca** | 124 | 16.4% |
| **Australia** | 5 | 0.5% | **Riyadh** | 254 | 33.6% |
| **Azerbaijan** | 4 | 0.4% | **Hail** | 31 | 4.1% |
| **Bahrain** | 27 | 2.5% | **Eastern Region** | 66 | 8.7% |
| **Canada** | 5 | 0.5% | **Asir** | 21 | 2.8% |
| **Egypt** | 48 | 4.5% | **Qassim** | 135 | 17.9% |
| **France** | 12 | 1.1% | **Taif** | 4 | 0.5% |
| **Germany** | 10 | 0.9% | **Jizan** | 6 | 0.8% |
| **Indonesia** | 10 | 0.9% | **Listed location unknown** | 8 | 1.1% |
| **Jordan** | 21 | 2.0% | **Bahah** | 10 | 1.3% |
| **Kuwait** | 25 | 2.4% | **Tobouk** | 32 | 4.2% |
| **Lebanon** | 5 | 0.5% | **Najran** | 6 | 0.8% |
| **Libya** | 14 | 1.3% | **Medina** | 26 | 3.4% |
| **Morocco** | 5 | 0.5% | **Unlisted** | 18 | 2.4% |
| **Palestine** | 5 | 0.5% | **N. Borders** | 6 | 0.8% |
| **Qatar** | 8 | 0.8% | **Jouf** | 8 | 1.1% |
| **Russia** | 3 | 0.3% | | | |
| **Saudi Arabia** | 755 | 71.3% | | | |
| **Sudan** | 4 | 0.4% | **Arabian Peninsula-Wide Sample (cont. from left)** | | |
| **Syria** | 7 | 0.7% | Afghanistan, Albania Belgium, Denmark, Finland, Iran, Kosovo, Macedonia, Malaysia, Norway, Pakistan, Romania, South Africa, Spain, Sweden, Tanzania, Uzbekistan. | All had 1 fighter | 1.7% (sum of all 17 countries) |
| **Tunisia** | 22 | 2.1% | China, Holland, India, Iraq, Oman. | All had 2 fighters | 2.0% (sum of all 5 countries) |
| **Turkey** | 5 | 0.5% | | | |
| **UAE** | 5 | 0.5% | | | |
| **United Kingdom** | 10 | 0.9% | | | |
| **USA** | 4 | 0.4% | | | |
| **Yemen** | 21 | 2.0% | | | |

## Appendix V – Data Accessibility and Ethics

### *Anonymization*

Subjects in the dataset have been anonymized by replacing their names and their chosen *nom de guerre* (which are not unique but could be used to identify the individual) with an ID number. I also re-coded home addresses (if provided) from specific locations to the largest, subnational geographic region (i.e., states or provinces). Finally, edge data such as the names of references or facilitators only appear as numbers in my adjacency matrices. Therefore, personally identifiable information about these individuals is also removed.

### *How the Data Were Obtained*

Individual consent was not obtained from the individuals and cannot be obtained from these individuals because they have joined a terrorist group. Instead, the individuals have been anonymized as described above.

In early 2016, an ISIS fighter defected with personnel records containing detailed registration forms recorded by ISIS border officials who interviewed over 3,500 newly arrived foreign recruits from 2013-2014. I was given these forms by a Syrian journalist in the spring of 2016. I verified the validity of these data by cross-referencing personal data in the registration files against research that I had overseen which uncovered the identities of ISIS fighters whose names were not public. I also verifying that my findings were consistent with other studies that had used the same dataset (e.g., Dodwell et al, 2016). Several other organizations also have copies of this same dataset, and I have discussed these data with them, including researchers at West Point Counter-Terrorism Center, George Washington University and the Royal United Services Institute (RUSI), based in London.

## Appendix VI - ERGMs of France and Saudi Networks to Assess Provincial-Level Ties

Saudi Arabia and France were chosen for the province-level analysis because they were the countries with the largest number of fighters in their respective regions (Saudi Arabia n=755, and France n=108). The findings from Table 6 below confirm that which was presented at the regional level: geography was a far more power predictor of fighter ties in a European country (France) than they are in a country in the Arabian Peninsula (Saudi Arabia).

Table 6 – Results of ERGMs for France and Saudi Arabia Networks

| | **Reference Network** | | **Facilitator Network** | | **Entry Network** | |
|---|---|---|---|---|---|---|
| | France | S. Arabia | France | S. Arabia | France | S. Arabia |
| ***Province of Residence*** | ***2.11 (0.2)****** | ***1.13 (0.06)****** | ***0.56 (0.15)****** | ***-0.04 (0.01)****** | ***0.65 (0.24)***** | ***0.86 (0.08)****** |
| Birth Year | 1.14 (0.27)*** | 0.5 (0.09)*** | 0.12 (0.23) | 0.03 (0.02) . | 0.00 (0.4) | 0.80 (0.1)*** |
| Nationality | 0.12 (0.2) | 0.5 (0.1)*** | -0.86 (0.15)*** | 0.42 (0.01)*** | 0.00 (.21) | 0.78 (0.15)*** |
| Education | 0.61 (0.2)** | 0.18 (0.06)** | 0.98 (0.11)*** | 0.01 (0.01) | 0.65 (0.2)** | 0.14 (0.08)· |
| Religious Knowledge | -0.12 (0.21) | 0.2 (0.06)*** | 0.31 (0.12)** | 0.46 (0.01)*** | 1.41 (0.26)*** | 0.40 (0.07)*** |
| Profession | 0.22 (0.22) | 0.12 (0.07)· | -0.42 (0.15)** | 0.1 (0.01)*** | -0.2 (0.24) | 0.6 (0.08)*** |
| Prev. Jihad Experience | 2.14 (0.4)*** | 0.23 (0.09)** | -0.58 (0.12)*** | 0.46 (0.01)*** | 0.45 (0.26) . | 0.78 (0.14)*** |
| Marital Status | 0.39 (0.21)· | 0.08 (0.06) | 0.1 (0.12) | 0.19 (0.01)*** | 0.3 (0.2) | 0.1 (0.08) |
| Countries Visited | 0.41 (0.22)· | 0.47 (0.06)*** | 0.47 (0.13)*** | 0.19 (0.01)*** | 0.59 (0.22)** | 0.65 (0.08)*** |
| *Edges* | *-6.93 (0.42)**** | *-6.78 (0.13)**** | *-2.87 (0.11)**** | *-1.81 (0.02)**** | *-5.76 (.3)**** | *-8.22 (0.2)**** |
| AIC/BIC | 925/991 | 15251/15356 | 2493/2559 | 330923/331029 | 1010/1076 | 394588/284635 |

*Standard errors in parenthesis; '***' $p < 0.001$, '**' $p < 0.01$, '*' $p < 0.05$, '.' $p < 0.1$; France (N=109); Saudi Arabia (N=755)*

Analyzing province-level attributes in France shows that shared provincial ties was the most significant predictor of ties between fighters from France. In the reference network, fighters from the same province in France were over eight times as likely to share a tie. Whereas in the facilitator network, fighters were less likely to share ties if they were from the same province than if they were from the same country, possibly due to a deliberate sorting of European fighters based on their country of origin (see "ISIS Managing Fighter Arrivals" below). In the facilitator network, the predictive effects of subnational-level residence for France were slightly stronger: a fighter was 1.3 times more likely to share a tie with another fighter from the same country while 1.7 times more likely to share a tie with another fighter if they were from the same province.

There are several other attributes in the network from France that also predict ties between fighters. Nationality is a strong predictor of ties in the reference network, but the effect is negative in the facilitator network and negligible in the entry network. Those who claim French residence and previous jihad experience are more likely to share a tie in the reference network than any other attribute, including province of origin.

On the other hand, fighters from Saudi Arabia were more likely to know each other if they were from the same province than any other attribute. That fighters from Saudi Arabia were more likely to know someone from their province and almost equally likely to be connected to a foreigner at the country level suggests a widespread and grassroots mobilization network.